\documentclass[twoside]{article}
\usepackage{fleqn,espcrc2,epsfig}

\newcommand{\newc}{\newcommand}
\newc{\hc}{{\it {h.c.}}}
\newc{\ie}{{\it  {i.e.}}}
\newc{\eg}{{\it  {e.g.}}}
\newc\etal{{\it {et al.}}}
\newc{\etc}{{\it {etc.}}}

\newcommand\lsim{\mathrel{\rlap{\lower4pt\hbox{\hskip1pt$\sim$}}
    \raise1pt\hbox{$<$}}}
\newcommand\gsim{\mathrel{\rlap{\lower4pt\hbox{\hskip1pt$\sim$}}
    \raise1pt\hbox{$>$}}}

\newc{\mx}{M_{GUT}}
\newc{\gx}{g_{GUT}}     \newc{\alphax}{\alpha_{\rm GUT}}

\newcommand\fa{f_{a}}
\newcommand\mchi{m_{\chi}}

\newc{\tanb}{\tan\beta}
\newc{\mtop}{m_t}
\newc{\mbot}{m_b}
\newc{\mtau}{m_{\tau}}
\newc{\htop}{h_t}
\newc{\hbot}{h_b}
\newc{\htau}{h_{\tau}}
\newc{\stopq}{{\widetilde t}}   
\newc{\mstop}{m_{\stopq}}  
\newc{\stp}{\stopq}
\newc{\stopl}{\widetilde t_L}
\newc{\stopr}{\widetilde t_R}
\newc{\stopone}{\widetilde t_1}
\newc{\stoptwo}{\widetilde t_2}

\newc{\mstopq}{m_{\tilde t}}
\newc{\mstopl}{m_{\tilde t_L}}
\newc{\mstopr}{m_{\tilde t_R}}
\newc{\mstopone}{m_{\tilde t_1}}
\newc{\mstoptwo}{m_{\tilde t_2}}
\newc{\stau}{{\widetilde \tau}}
\newc{\staul}{\widetilde \tau_L}
\newc{\staur}{\widetilde \tau_R}
\newc{\stauone}{\widetilde \tau_1}
\newc{\stautwo}{\widetilde \tau_2}

\newc{\azero}{A_0}
\newc{\bzero}{B_0}
\newc{\muzero}{\mu_0}
\newc{\sgnmu}{{\rm sgn}\,\mu}

\newcommand\photino{\widetilde{\gamma}} \newcommand\mphotino{m_{\photino}}

\newcommand\axino{\widetilde{a}}        \newcommand\maxino{m_{\axino}}
\newcommand\abunda{\Omega_{\axino}h^2}

\newc{\cachigamma}{C_{a\chi\gamma}}

\newc{\caww}{C_{aWW}}                   \newc{\cayy}{C_{aYY}}
\newc{\sthw}{\sin\theta_W}              \newc{\cthw}{\cos\theta_W}
\newc{\bino}{\widetilde B}              \newc{\wino}{\widetilde W_3}
\newc{\higgsinob}{{\widetilde H}^0_b}   \newc{\higgsinot}{{\widetilde H}^0_t}
\newc{\mcharone}{m_{\charone}}  \newc{\charone}{\chi_1^\pm}
\newc{\sfermion}{\widetilde f}
\newc{\msf}{m_{\sfermion}}

\newc{\abund}{\Omega h^2}
\newc{\omegachi}{\Omega_\chi}
\newc{\abundchi}{\Omega_\chi h^2}
\newc{\rhocrit}{\rho_{crit}}
\newc{\rhochi}{\rho_{\chi}}

\newc{\sigmaann}{\sigma_{\rm ann}}

\newc{\sigmap}{\sigma_p}

\newc{\mwimp}{m_{\rm WIMP}}     \newc{\rhowimp}{\rho_{\rm WIMP}}

\newc{\mplanck}{M_{\rm Planck}}              \newc{\mgut}{M_{\rm GUT}}
\newc{\mz}{m_{Z}}                       \newc{\mw}{m_{W}}

\newc{\xf}{x_f}
\newc{\jxf}{J({\xf})}
\newc{\VEV}[1]{\langle #1 \rangle}

\newcommand\tev{\,\mbox{TeV}}
\newcommand\gev{\,\mbox{GeV}}
\newcommand\mev{\,\mbox{MeV}}
\newcommand\kev{\,\mbox{keV}}
\newcommand\ev{\,\mbox{eV}}

\newc{\ra}{\rightarrow}
\newc{\beq}{\begin{equation}}
\newc{\eeq}{\end{equation}}
\newc{\be}{\begin{equation}}
\newc{\ee}{\end{equation}}

\newc{\ekd}{\; {\rm event}/kg/{\rm day}}
\newc{\ekds}{\; {\rm events}/kg/{\rm day}}
\newc{\ekkds}{\; {\rm events}/kg/keV/{\rm day}}
\newc{\kgday}{\;\mbox{kg$\times$day}}
\newc{\pb}{\,\mbox{pb}}
\newc{\rhozerothree}{\rho_{0.3}}
\newc{\gevcmcube}{\,\mbox{GeV/cm$^3$}}

\newc{\sigchin}{\sigma(\chi N)}

\def\PLB#1#2#3{Phys. Lett. B {\bf#1} (19#2) #3}

\def\PRD#1#2#3{Phys. Rev. D {\bf#1} (19#2) #3}

\begin{document}

\begin{titlepage}
\pagestyle{empty}
\baselineskip=21pt
\rightline{CERN--TH/2001-061}
\vskip 0.5in
\begin{center}
{\large {\bf Non-Baryonic Dark Matter\footnote{Invited plenary 
review talk given at 6th International Workshop on Topics in 
Astroparticle 
and Underground Physics (TAUP 99), 6-10 September, 1999, Paris,
France.}
}}
\end{center}
\begin{center}
\vskip 0.05in
{\bf Leszek Roszkowski}$^{1,2}$\\

\vskip 0.05in
{\it
$^1${Department of Physics, Lancaster University, Lancaster LA1
4YB, England}\\
$^2${TH Division, CERN, CH-1211 Geneva 23, Switzerland}\\
}
\vskip 0.5in
{\bf Abstract}
\end{center}
\baselineskip=18pt 
\noindent 
There exist several well-motivated candidates for non-baryonic cold
dark matter, including neutralinos, axions, axinos, gravitinos,
Wimpzillas.  I review the dark matter properties of the neutralino
and the current status of its detection. I also discuss the axino as
a new interesting alternative.

\vfill
\vskip 0.15in
\leftline{CERN--TH/2001-061}
\leftline{February 2001}
\end{titlepage}

\newpage

\title{Non-Baryonic Dark Matter
\thanks{Invited plenary 
review talk given at 6th International Workshop on Topics in 
Astroparticle 
and Underground Physics (TAUP 99), 6-10 September, 1999, Paris,
France.}
}

\author{Leszek Roszkowski\\
\vspace{1pc}        
Department of Physics, Lancaster University,\\ 
Lancaster LA1 4YB, England}

\begin{abstract}
  
There exist several well-motivated candidates for non-baryonic cold
dark matter, including neutralinos, axions, axinos, gravitinos,
Wimpzillas.  I review the dark matter properties of the neutralino
and the current status of its detection. I also discuss the axino as
a new interesting alternative.

\vspace{1pc}
\end{abstract}


\maketitle

\section{Introduction}

The puzzle of the hypothetical dark matter (DM) in the Universe still
remains unresolved~\cite{kt}. 
While there may well be more than one
type of DM, arguments from large structures suggest that a
large, and presumably dominant, fraction of DM in the Universe is made
of  massive particles which at the time of entering the epoch of matter
dominance would be already non-relativistic, or {\em cold}.  From the
particle physics point of view, cold DM (CDM) could most plausibly be
made of so-called weakly interacting massive particles (WIMPs).

There exist several interesting WIMP candidates for CDM that are
well-motivated by the underlying particle physics. The neutralino is
considered by many a ``front-runner'' by being perhaps the most
natural WIMP: it comes as an unavoidable prediction of supersymmetry
(SUSY).  The axion is another well-motivated
candidate~\cite{sikivie:taup}.  But by no means should one forget
about some other contenders. While some old picks (sneutrinos and
neutrinos with mass in the GeV range) have now been ruled out as
cosmologically relevant CDM by LEP, axinos (SUSY partners of axions)
have recently been revamped~\cite{ckr}. Another well-known candidate
is the fermionic partner of the graviton, the gravitino.  Most of the
review will be devoted to the neutralino but I will also comment about
recent results regarding axinos.

Neutrinos, the only WIMPs that are actually known to exist, do not
seem particularly attractive as DM candidates. It has long been
believed that their mass is probably very tiny, as suggested by
favored solutions to the solar and atmospheric neutrino problems,
which would make them hot, rather than cold DM. This picture has
recently been given strong support by first direct evidence from
Superkamiokande for neutrinos' mass~\cite{superkamiokande}. While the
new data only gives the $\mu-\tau$ neutrino mass-square difference of
$2.2\times10^{-3}\ev^2$, it it very
unlikely that there would exist two massive neutrinos with
cosmologically relevant mass of  5 to 40 eV and such a tiny mass
difference.

The strength with which the WIMP candidates listed above interact with
ordinary matter spans many orders of magnitude. For the neutralinos it
is a fraction ($\sim 10^{-2} - \sim 10^{-5}$) of the weak
strength. Interactions of axions and axinos are suppressed by
$(m_W/\fa)^2\sim 10^{-16}$, where $\fa\sim10^{10}\gev$ is the scale at
which the global Peccei-Quinn $U(1)$ symmetry is broken. Interactions
of gravitinos and other relics with only gravitational interactions
are typically  
suppressed by $(m_W/\mplanck)^2\sim 10^{-33}$. One may wonder how such
vastly different strengths can all give the relic abundance of the
expected order of the critical density. The answer lies in the
different ways they could be produced in the early Universe: the
neutralinos are mainly produced ``thermally'': at some temperature they
decouple from the thermal bath. But in addition there exist also
several mechanisms of ``non-thermal'' production. This is how CDM
axinos, gravitinos can be produced (in addition to ``thermal'' production), 
as well as Wimpzillas~\cite{wimpzilla}.

It is obvious that WIMPs do not necessarily have to interact only via
weak interactions {\em per se}. WIMPs are generally required to be
electrically neutral because of stringent observational constraints on
the abundance of stable charged relics.  On the other hand, they
could in principle carry color charges. (For example, a stable gluino
above $130-150\gev$~\cite{gunion}, or in an experimentally allowed
window $25 - 35\gev$~\cite{raby}, could still be the lightest SUSY
particle (LSP) although its
relic abundance would be very small~\cite{gunion}.) In a halo they
would exist as neutral states by forming composites with gluon or
quarks.  We can see that generally WIMPs are expected to have
suppressed effective couplings with ordinary matter. Otherwise, they
would dissipate their kinetic energy.

What is the WIMP relic abundance
$\abundchi=\rhochi/\rhocrit$~\cite{primack:taup}?
Current estimates of the lower bound on the age of the Universe lead
to $\Omega_{\rm TOTAL} h^2<0.25$.  Recent results from rich clusters of
galaxies and high-redshift
supernovae type 
Ia imply $\Omega_{\rm matter}\simeq0.3$. The Hubble parameter is now
constrained to $0.65\pm0.1$. Big Bang nucleosynthesis requires
$\Omega_{\rm baryon} h^2\lsim0.015$. Assuming that most matter in the
Universe is made of CDM WIMPs, one therefore obtains
\beq
0.1\lsim \abundchi\lsim0.15
\eeq
or so. 
At the very least, requiring that a dominant fraction of CDM is
located only in galactic halos gives $\abundchi\gsim0.025$.


\section{Neutralinos}

The DM candidate that has attracted perhaps the most wide-spread
attention from both the theoretical and experimental communities is
the neutralino. It is a neutral Majorana particle, the lightest of the
mass eigenstates of the fermionic partners of the gauge and Higgs
bosons: the bino, wino and the two neutral higgsinos. If it is the
lightest SUSY particle, it is massive and stable, due to assumed
R-parity. A perfect candidate for a WIMP!

The neutralino is a well-motivated candidate. It is an inherent
element of any phenomenologically relevant SUSY model. Being neutral,
it is a natural candidate for the LSP (although one should remember
that this is often only an assumption); it couples to ordinary matter
with a weak-interaction strength (reduced by mixing angles) which is
within the range of sensitivity of present-day high energy, as well as
dark matter, detectors. Finally, as a bonus, it naturally gives
$\abundchi\sim1$ for broad ranges of masses of SUSY particles below a
few \tev\ and for natural ranges of other SUSY parameters.

Much literature has been devoted to the neutralino as DM, including a
number of comprehensive and topical reviews (see, \eg,
Refs.~\cite{jkg,vietnam95}). Here I will only summarize the
main results and comment on some recent developments and
updates.

Neutralino properties as DM and ensuing implications for SUSY spectra
are quite model dependent but certain general conclusions can be
drawn. First, its cosmological properties are very different depending
on a neutralino type. The relic abundance of gaugino-like (mostly
bino-like) neutralinos
is primarily determined by the (lightest) sfermion exchange in the LSP
annihilation into $f\bar f$: $\abund\propto \msf^4/\mchi^2$. In order
to have $\omegachi\sim1$, the lightest sfermion cannot be too light
(below $\sim100\gev$) nor too heavy (heavier than a few hundred
\gev)~\cite{chiasdm} which is a perfectly natural range of values. 

On the other hand, higgsino-like $\chi$'s are
strongly disfavored. Firstly, due to GUT relations among gaugino
masses, higgsinos correspond to rather large gluino mass values, above
1~\tev, which may be considered as unnatural~\cite{chiasdm}. 
Furthermore, higgsino-like neutralinos have been shown to provide very
little relic abundance.
For $\mchi>\mz,\mw,\mtop$ the $\chi$ pair-annihilation into
those respective final states ($ZZ$, $WW$, $t\bar t$) is very
strong.  But both below and above those thresholds, there are
additional co-annihilation \cite{coann} processes of the LSP with
$\charone$ and $\chi^0_2$, which are in this case almost
mass-degenerate with the LSP. Co-annihilation typically reduces $\abundchi$
below any interesting level although it has been recently argued that
the effect of co-annihilation is not always as strong as previously
claimed~\cite{coann-new}.
Higgsino-like LSPs are thus rather unattractive, although still possible 
DM candidates, especially in the large mass ($\mchi\gsim500\gev$) regime.
One also arrives at the same conclusions in the case of
`mixed' neutralinos composed of comparable fractions of gauginos and
higgsinos.  This is because, even without co-annihilation, in this
case the neutralino pair-annihilation is less suppressed and one
invariably finds very small $\omegachi$ there~\cite{chiasdm}.

Remarkably, just such a cosmologically preferred gaugino type of
neutralino typically {\em emerges} in a grand-unified scenario with
additional assumptions that the mass parameters of all spin-zero
particles are equal at the unification scale $\sim10^{16}\gev$. What
one finds there is that the lightest bino-like neutralino emerges as
essentially the only choice for a neutral LSP~\cite{dick,kkrw}. (It is
still possible to find cases with a higgsino-like LSPs but they are
relatively rare.)  Furthermore, in order for $\abundchi$ not to exceed
one or so, other (most notably sfermion) SUSY partner masses have to
be typically less than 1~\tev~\cite{dick,kkrw}. Thus our
phenomenological expectations for low-energy SUSY to be realized
roughly below 1~\tev\ are nicely consistent with a bino-like
neutralino as a dominant component of DM in the Universe.

What about the neutralino mass? In principle, it could be considered
as a nearly free parameter. It would be constrained from below by the
requirement $\abundchi<1$ to lie above some $2 - 3~\gev$~\cite{gr92}
which is a neutralino version of the Lee-Weinberg
bound~\cite{kt}. This is because, in this mass range, the neutralino
may be viewed as a massive neutrino with somewhat suppressed
couplings. Much stronger bounds are in many cases provided by
LEP. Unfortunately, collider bounds are rather model dependent and are
no longer provided for the general SUSY model normally considered in
DM studies. Nevertheless, one can reasonably expect that $\mchi$ is now
ruled out below roughly 30\gev, except in cases when the sneutrino is
nearly degenerate with the LSP in which case the lower bound may
disappear altogether.

A rough upper bound of 1~\tev\ follows from the theoretical criterion
of naturalness: expecting that no SUSY masses should significantly
exceed that value. Again, GUT relations among gaugino masses cause the
above constraint to provide a much more stringent (although still only
indicative) bound of about $150 - 200~\gev$. Remarkably, in the
scenario with additional GUT unification of spin-zero mass parameters,
just such a typical upper bound derives from the cosmological
constraint $\abundchi<1$. Only in a relatively narrow range of
parameters where the neutralino becomes nearly mass degenerate with a
SUSY partner of the tau, $\widetilde\tau_R$, the effect of the
neutralino's co-annihilation opens up the allowed range of $\mchi$ up
to about $600\gev$~\cite{efos99-coann}.

Summarizing, we can see that, despite the complexity of the neutralino
parameter space and a large number of neutralino annihilation
channels, one can, remarkably, select {\em the gaugino-like neutralino in
the mass range between roughly 30 and 150\gev\ as a natural and attractive
DM candidate}. Furthermore, one is able to derive relatively stringent
conditions on the mass range of some sfermions, which are consistent
with our basic expectations for where SUSY might be realized.

\section{Neutralino WIMP Detection}\label{detection:sec}

\subsection{Predictions}\label{predictions:sec}

The local halo density of our  Milky Way is estimated at 
$0.3\gev/{\rm cm^3}$ with a factor of two or so
uncertainty~\cite{jkg}. For neutralino WIMPs as a dominant component
of the halo this translates to about 3000 LSPs with mass $\mchi=100\gev$ per
cubic meter. With typical velocities in the range of a few hundred
$km/s$, the resulting flux of WIMPs is actually quite large 
\beq
\Phi=v\rho_\chi/\mchi\approx10^{9}\left(100\gev/\mchi\right) {\rm
\chi's}/m^2/s.  
\eeq

A massive experimental WIMP search programme has been developed during
the last few years. Although a variety of techniques have been
explored, most of them follow one of the two basic strategies.  One
can look for DM neutralinos {\em directly}, through the halo WIMP
elastic scattering off nuclei, $\chi N\ra \chi N$, in a detector. {\em
Indirect} searches look for traces of decays of WIMP pair-annihilation
products. One promising way is to look for multi-GeV energy neutrinos
coming from the Sun and/or the core of the Earth. One can also look
for monochromatic photons, positrons or antiprotons produced in WIMP
pair-annihilation in a Galactic halo. I will not discuss these
indirect methods here. Perhaps I should only mention about an
interesting new way of looking for WIMPs at the Galactic
center~\cite{gondolosilk99}. If there is a super-massive
black hole there (for which there is now some evidence), it will
accrete WIMPs and thus increase their density in the core. They will
then be annihilating much more effectively and the resulting flux of
neutrinos, photons and other products from the center of the Milky Way
may be strongly enhanced, even up to a factor of $10^5$ in halo models
with central spikes in their profiles. Such spiked halo models have
been obtained in recent N-body simulations~\cite{moore}.

In the following I would like to make several comments about direct
detection and in particular about an intriguing claim made by the DAMA
Collaboration regarding a possible evidence for a WIMP signal in their
data. For a  more detailed review, see Ref.~\cite{morales}.
First, I will briefly summarize the theoretical aspects and predictions.

In direct searches one of the most significant quantities is the event
rate $R\sim \sigchin\, (\rho_\chi/\mchi v)\,(1/m_N)$ - the product
of the elastic cross section $\sigma(\chi N)$ of neutralinos from
nuclei, their flux $\rho_\chi/\mchi v$ and the density of target nuclei with
mass $m_N$. 

The elastic cross section $\sigchin$ of relic WIMPs scattering
off nuclei in the detector depends on the individual cross sections of
the WIMP scattering off constituent quarks and gluons.  For
non-relativistic Majorana particles, these can be divided into two
separate types. The coherent part described by an effective
scalar coupling between the WIMP and the nucleus is proportional to
the number of nucleons in the nucleus. It receives a tree-level
contribution from scattering off quarks, $\chi q\ra \chi q$, as
described by a Lagrangian ${\cal L}\sim
\left(\chi\chi\right)\left(\bar q q\right)$.  The incoherent component
of the WIMP-nucleus cross section results from an axial current
interaction of a WIMP with constituent quarks, given by ${\cal L}\sim
\left(\chi\gamma^\mu\gamma_5\chi\right)\left(\bar q\gamma_\mu\gamma_5
q\right)$, and couples the spin of the WIMP to the total spin of the
nucleus.

The differential cross section for a WIMP scattering
off a nucleus $X_{Z}^{A}$ with mass $m_A$ is therefore given by
\begin{eqnarray}
\frac{d\sigma}{d|\vec{q}|^2}=\frac{d\sigma^{scalar}}{d|\vec{q}|^2}+
\frac{\ d\sigma^{axial\ }}{d|\vec{q}|^2},
\label{signucleus:eq}
\end{eqnarray}
where the transferred momentum $\vec{q}=\frac{m_A
\mchi}{m_A+\mchi}\vec{v}$ depends on the velocity $\vec{v}$ of the
incident WIMP.  The effective WIMP-nucleon cross sections
$\sigma^{scalar}$ and $\sigma^{axial}$ are computed by evaluating
nucleonic matrix elements of corresponding WIMP-quark and WIMP-gluon
interaction operators.  

In the scalar part contributions from individual
nucleons in the nucleus add coherently and the finite size effects are
accounted for by including the scalar nuclear form factor $F(q)$.
(The effective interaction in general also
includes tensor components but the relevant nucleonic matrix elements
can be expanded in the low momentum-transfer limit in terms of the
nucleon four-momentum and the quark (gluon) parton distribution
function.  As a result, the non-relativistic WIMP-nucleon Lagrangian
contains only scalar interaction terms.) 
The differential cross section for the scalar part then 
takes the form \cite{jkg}     
\begin{eqnarray}
\frac{d\sigma^{scalar}}{d|\vec{q}|^2}=\frac{1}{\pi v^2}[Z f_p +(A-Z) f_n]^2
F^2 (q),
\end{eqnarray}
where $f_{p}$ and $f_{n}$ are the effective WIMP couplings to
protons and neutrons, respectively, and typically 
$f_{n}\approx f_{p}$. Explicit expressions for the case
of the supersymmetric neutralino can be found, \eg, in the Appendix of 
Ref.~\cite{bb97}.

The effective axial
WIMP  coupling to the nucleus  depends on the spin content
of the nucleon $\Delta q_{p,n}$ and the overall expectation value of the 
nucleon group spin in the nucleus $<S_{p,n}>$. For a nucleus with a total
angular momentum $J$ we have
\begin{eqnarray}
\frac{d\sigma^{axial}}{d|\vec{q}|^2}=\frac{8}{\pi v^2}\Lambda^2 J (J+1)
S(q),
\end{eqnarray}
with $\Lambda=\frac{1}{J} [a_p \langle S_p\rangle+a_n \langle S_n\rangle]$.
The axial couplings 
\beq
a_p=\frac{1}{\sqrt{2}} \sum_{\scriptstyle u,d,s} d_q \Delta q^{(p)},\
\ \ 
a_n=\frac{1}{\sqrt{2}} \sum_{\scriptstyle u,d,s} d_q \Delta q^{(n)}
\eeq
are determined by the experimental values of the spin constants $\Delta 
u^{(p)}=\Delta d^{(n)}=0.78$, $\Delta d^{(p)}=\Delta u^{(n)}=-0.5$ and 
$\Delta s^{(p)}=
\Delta s^{(n)}=-0.16$.
The effective couplings $d_q$ depend on the WIMP properties and for
the neutralino they can be found, \eg, in the Appendix of Ref.~\cite{bb97}.

In translating $\sigma(\chi q)$ and $\sigma(\chi g)$ into the
WIMP-nucleon cross section in~Eq.~(\ref{signucleus:eq}) several
uncertainties arise. The nucleonic matrix element coefficients for the scalar
interaction are not precisely known. Also, the spin content of the nucleon
and the expectation values of the proton (neutron) group spin in a particular 
nucleus are fraught with significant uncertainty and nuclear model 
dependence. These ambiguities have to be considered in numerical calculations.
Finally, in order to obtain $\sigma(\chi N)$, models of nuclear
wave functions must be used. The scalar nuclear
form factor reflects  the mass density distribution in the
nucleus~\cite{jkg}.

The resulting cross section for scalar (or coherent) interactions is
\be
\sigma^{scalar}(\chi N) \sim G_F^2  
{{\mchi^2 m_N^2}\over{(m_N +\mchi)^2}} A^2
\label{sigma-coherent:eq}
\ee
and is proportional to the mass of the nucleus.

For axial (or incoherent) interactions one finds
\be
\sigma^{axial}(\chi N) \sim G_F^2 {{\mchi^2 m_N^2}\over{(m_N +\mchi)^2}}
\label{sigma-incoherent:eq}
\ee
which can be shown to be proportional to the spin of the
nucleus. ($G_F$ is the Fermi constant.)

Unfortunately, in supersymmetric models actual calculations produce a
rather broad range of values, $R\sim 10^{-5} - 10\ekds$. The rates are also
very small.
The reason why this is so is clear: this is because of
smallness of $\sigma(\chi N)$. The elastic cross section is related by
crossing symmetry to the neutralino annihilation cross section
$\sigmaann$ which is of weak interaction strength and such as to give
$\abundchi\sim1$, and therefore very small.

Such small event rates are clearly
an enormous challenge to experimentalists aiming to search for dark
matter. One may realistically expect that continuing SUSY searches in
high energy accelerators and improving measurements of $\Omega_{\rm
CDM}$ and the Hubble parameter 
will cause those broad ranges of $R$ to gradually
shrink. 

The not so good news is that the choices of SUSY parameters for which
one finds the favored range of the relic abundance,
$0.1\lsim\abundchi\lsim 0.15$, correspond to rather low values of the
event rate $R\lsim10^{-2}\ekds$ or so, typically about an order of
magnitude below the reach of today's detectors.

\subsection{Annual Modulation}\label{seasonal:sec}

One promising way of detecting a WIMP is to look for
yearly time variation in the measured energy spectrum. 
It has been pointed out \cite{dfs86,ffg88} that a halo WIMP signal
should show a periodic effect due to the Sun's motion through the
Galactic halo, combined with the Earth's rotation around the Sun. The
peaks of the effect are on the 2nd of June and half a year later.

The effect, called ``annual modulation'', would provide a convincing
halo WIMP signal. Unfortunately, in SUSY models the effect is usually
small, $\Delta R\lsim5\%$~\cite{jkg,bb97}.  With the absolute event rates
being already very small, it is going to be a great challenge to
detect the effect.

Here I would like to make some comments about possible evidence for a
WIMP signal in annual modulation that has been claimed by the DAMA
Collaboration.  Based on the statistics of $14,962\kgday$ of data
collected in a NaI detector over a period from November '96 to July
'97 (part of run~II), the Collaboration has reported \cite{dama98two}
a statistically significant effect which could be caused by an annual
modulation signal due to a WIMP with mass $\mchi$ and WIMP-proton
cross section $\sigmap$ given as 
\begin{eqnarray}
\mchi &=& 59\gev^{+22}_{-14}\gev,\\
\xi_{0.3}\sigmap &=& 
7.0^{+0.4}_{- 1.7}\times10^{-6}\pb
\label{damarange:eq}
\end{eqnarray}
at $99.6\%$~CL, 
where $\xi_{0.3}=\rhochi/\rhozerothree$ stands for the local WIMP mass
density $\rhochi$ normalized to $\rhozerothree=0.3\gevcmcube$. (See
also Figure~6 in Ref.~\cite{dama98two} for a $2\sigma$ signal region in the
($\mchi,\xi_{0.3}\sigmap$) plane.)
According to DAMA, the new analysis is consistent with and confirms
the Collaboration's earlier hint \cite{dama98one} for the presence of the
signal based on $4,549\kgday$ of data. 

The claimed effect comes from a few lowest bins of the scintillation
light energy, just above the software threshold of $2~\kev$, and
predominantly from the first bin ($2 - 3~\kev$). This is indeed what
{\em in principle} one should expect from the annual modulation
effect.  DAMA appears confident about the presence of the effect in
their data, and claims to have ruled out other
possible explanations, like temperature effects, radon contamination
or nitrogen impurities. According to DAMA, the effect is caused by
single hit events
(characteristic of WIMPs unlike neutron or gamma background)
with proper modulation of about one year, peak around June, and small
enough amplitude of the time dependent part of the
signal. 

Nevertheless, several experimental questions remain and cast much
doubt on the validity of the claim. Here I will quote some
which I find particularly important to clarify. First, as stated
above, the claimed effect comes from the lowest one or two energy
bins. This is indeed what one should expect from an annual modulation
signal. But is the effect caused by just one or two energy bins
statistically significant? This is especially important in light of
the fact that the shape of the differential energy spectrum $dR/dE$ in
the crucial lowest energy bins as measured by DAMA is very different
from the one measured by Gerbier, \etal~\cite{gerbierastropone} for
the same detector material (NaI). In Ref.~\cite{gerbierastropone} the
corrected-for-efficiency $dR/dE$ is about $10\ekkds$ at 3\kev,
decreasing monotonically down to about $2\ekkds$ at 6\kev. (See
Fig.~15 in Ref.~\cite{gerbierastropone}.)  In contrast, DAMA's spectrum
shows a dip down to $1\ekkds$ at $2\kev$, above which it increases to
nearly $2\ekkds$ at $4\kev$~\cite{dama98two,gaitskell:taup}. It is
absolutely essential for the controversy of the shape of $dR/dE$ in
the lowest energy bins to be resolved. Furthermore, DAMA's data from
the second run ($\sim15000\kgday$) shows that the background in the
crucial lowest bin ([2,3]\kev) is only about half or less of that in
the next bins~\cite{gaitskell:taup}. One may wonder why this would be the case.
Examining more closely the data in the constituent nine NaI crystals,
one finds a rather big spread in the event rates~\cite{gaitskell:taup}. In
detector~8, in the lowest energy bin one finds no contribution from the
background whatsoever! 

Two other groups which have also used NaI have
reported~\cite{gerbierastropone,ukdmcanomevents} robust evidence of
events of unexpected characteristics and unknown origin. The data of
both teams has been analyzed using a pulse shape analysis (PSA).
A small but statistically significant component
was found with the decay time even shorter than the one expected from
WIMPs or neutrons. While the population of those events appears to be
too small to explain DAMA's effect, a question remains not only about
their origin (contamination?, external effect?)  but also how they
contribute to the energy spectrum in the crucial lowest bins. DAMA
claims not to have seen such events.

DAMA has not yet published the data over a full annual
cycle.\footnote{See Note Added.} 
In particular, no evidence has been published of the signal
going down with time. So far, the claimed signal has been based on two
periods of data taking. Ref.~\cite{dama98one} was based on two
relatively short runs, one in winter and one in summer. The second
analysis~\cite{dama98two} used the data collected between
November and July. Much more data has been collected and will soon be
published.  One should hope that a full and clear analysis of statistical
and systematic errors will also be performed. 

In Ref.~\cite{abriola98} annual modulation was reanalyzed for
germanium and NaI. It was concluded that the effect would be too small
to be seen with current sensitivity. Particularly illuminating is
Fig.~6.a where DAMA's data from Ref.~\cite{dama98one} (run one) was
re-plotted along with an expected signal for the modulated part of the
spectrum for the central values of the ranges of the WIMP mass and
cross section ($\sigmap$) selected by DAMA. One can hardly see any
correlation between the data and the expected signal.

A controversy over DAMA's alleged signal is of experimental
nature. One may therefore hope that it will be definitively
resolved. It would be of particular importance for another experiment
using different detector material to put DAMA's claim to test.  The
CDMS cryogenic detection experiment using germanium and silicon
crystals at Stanford has now reached an adequate sensitivity and has
already ruled out about a half of the region selected by
DAMA~\cite{gaitskell:taup}.\footnote{See Note Added.} Let us
hope that DAMA's claim will be falsified soon.

\begin{figure}[t!]      
\centerline{\epsfig{file=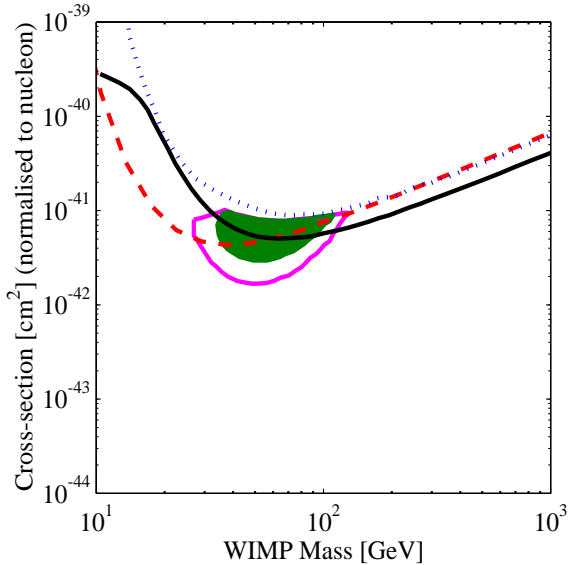,height=3in,width=3in}}
\bigskip
\caption{\small 
%
%
%
%
The current and some recent limits on the WIMP-proton spin-independent
(scalar) cross-section. The legend is as follows: 
dot: Heidelberg-Moscow Ge~\cite{hmlimit}, solid: DAMA
NaI~\cite{damalimit}, dash: CDMS 1999 $4\kgday$ Ge with 
neutron background 
subtraction~\cite{gaitskell:taup}. Marked also
are the $2-\sigma$ DAMA regions: solid closed curve: $15,000\kgday$,
filled region: 
$20,000\kgday$~\cite{dama98two}.
}
\label{status:fig}
\end{figure}
Figure~\ref{status:fig} shows the current status of  WIMP
searches in the plane of the WIMP-proton scalar cross-section
$\sigmap$ versus the WIMP mass.\footnote{The figure has been produced
with the help of a particularly useful plotting facility of Gaitskell
\&\ Mandic which is now available on the Web: 
http://cdms.berkeley.edu/limitplots/.} 
A claimed
DAMA region is also indicated.

It is worth pointing out that the claimed signal region selected by
DAMA~\cite{dama98two} was too restrictive as it did not include
astrophysical uncertainties. The effect is rather sensitive to
assumptions about a model of the Galactic halo.  In the DAMA
analysis the peak of the WIMP (Maxwellian) velocity distribution was
fixed at $v_0=220\, km/s$.  Since the Galactic halo has not been
directly observed and there still is a considerable disagreement about
a correct halo model, quoted error bars for $v_0$ and the local
halo density should, in my opinion, be treated with much caution.
Varying $v_0$ within a reasonable range leads to a significant
enlargement of the region selected by DAMA, \cite{dama98two} as first
shown in Ref.~\cite{br99}. This is so because of the significant
dependence on $v_0$ of the differential event rate spectrum, as can be
seen in Fig.~1 in Ref.~\cite{br99}.

As a result, the upper limit of $\mchi$ corresponding to the claimed
signal region increases considerably, from about $100\gev$ for the
initially assumed value $v_0=220\, km/s$ up to over $180\gev$ at
$v_0=170\, km/s$ \cite{damav99} (at $2\sigma$~CL). On the other hand
the range of $\xi_{0.3}\sigma_p$ is not much affected.

\begin{figure}[t!]      
\centerline{\epsfig{file=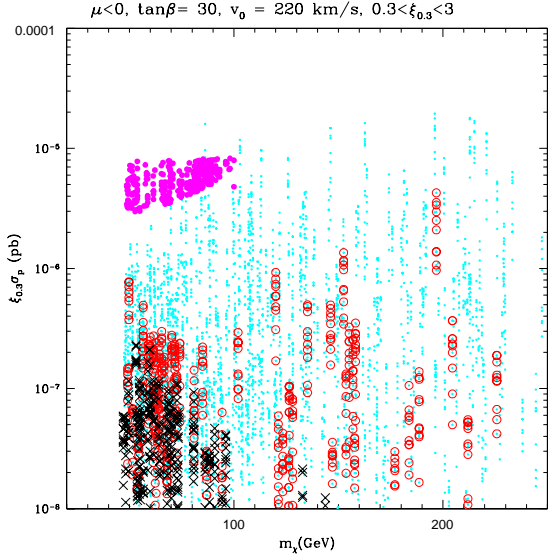,height=3in,width=3in}}
\bigskip
\caption{\small 
%
%
%
%
A scan of SUSY points in the plane ($\mchi,\xi_{0.3}\sigma_p$) where
$\xi_{0.3}= \rhochi/(0.3\gev/cm^3)$. The legend: circles:
$0.1<\abundchi<0.15$ (cosmologically favored range), crosses:
$\abundchi>0.25$ (excluded), small points: $0.02<\abundchi<0.25$ 
(conservative range).
The $2-\sigma$ region (upper left) of DAMA is marked with full thick dots. 
}
\label{damavssusy:fig}
\end{figure}

Assuming that the effect claimed by DAMA were indeed caused by DM
WIMPs, it is interesting to ask what ranges of SUSY parameters and LSP
relic abundance it would correspond to. This issue was addressed by
three groups, each using their own code for computing the direct
detection rate and the relic abundance~\cite{bdfs,na98,bgr99}. While
there is some difference in approach (somewhat different ranges of
input parameters, techniques and assumptions, \etc) the overall
outcome is that it is indeed possible to find such SUSY configurations
which could reproduce the signal but only for large enough $\tan\beta$
(typically above 10 although SUSY points with smaller values can also
sometimes be found).  More importantly, the corresponding values of
$\abundchi$ are rather small, typically below
0.06~\cite{bdfs,na98,bgr99} although somewhat larger values can also
sometimes be found.  These are clearly small values, well below the
favored range $0.1\lsim\abundchi\lsim 0.15$.  This is illustrated in
Figure~\ref{damavssusy:fig}~\cite{bgr99}.

However, one should remember that these results have been obtained
using commonly used but often rather uncertain values of several input
parameters. In addition to astrophysical ones, the nuclear physics and
quark mass inputs in calculating the scalar
cross-section for the neutralino-nucleus elastic scattering are rather
poorly determined, as mentioned above. The effect of the latter has
been recently re-analyzed in Ref.~\cite{bdfs99} and found to affect
the overall scalar cross-section by a factor of ten or so.

\begin{figure}[t!]      
\centerline{\epsfig{file=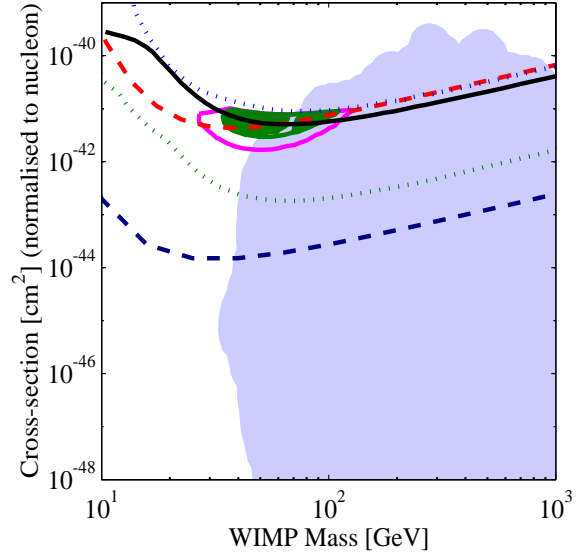,height=3in,width=3in}}
\bigskip
\caption{\small 
The current and some future reach limits on the WIMP-proton
spin-independent (scalar) cross-section compared with conservative
predictions from minimal SUSY (large shaded region). 
The legend is as follows: upper dot: Heidelberg-Moscow
Ge~\cite{hmlimit}, 
solid: DAMA NaI~\cite{damalimit}, upper dash: CDMS 1999 $4\kgday$ Ge
with neutron
background subtraction~\cite{gaitskell:taup}. Marked also are the
$2-\sigma$ DAMA regions: solid: $15,000\kgday$, filled:
$20,000\kgday$~\cite{dama98two}. Future reach of some experiments (as
claimed by the respective groups): lower dash: CDMS at Soudan, lower 
dot: CRESST. Some other experiments (\eg, Edelweiss,
UKDMC-Xe, GENIUS) expect to reach roughly similar
sensitivities. (Source: http://cdms.berkeley.edu/limitplots/).
}
\label{future:fig}
\end{figure}

It is worth presenting Figure~\ref{future:fig} again with an outlook for some
expected limits (as claimed by the respective groups) and compare them
with predictions of minimal SUSY obtained for very broad ranges of
SUSY parameters and neglecting nuclear input
uncertainties mentioned above. The SUSY region is bounded from above
by a somewhat arbitrary requirement $\abundchi>0.025$. From below it
is limited
by a generous bound $\abundchi<1$. The SUSY points
falling into the currently expected range
$0.1\lsim\abundchi\lsim0.15$ (not indicated in the plane)
form a sub-region reaching up to roughly a few times
$10^{-6}$~pb. 

It is clear that today's experiments are now only reaching the
sensitivity required to begin testing predictions coming from
SUSY. What I find promising is that several experiments using
different detector materials and often different methods of [attempts
at] distinguishing signal from background will explore a large
fraction of the SUSY parameter space within the next few years. 
Especially reassuring would be an observation of a positive signal in more
than type of DM detector, although many experimentalists would probably
remark that I am asking for too much. Time will tell.

\section{Axinos}

Axinos are a natural prediction of the Peccei-Quinn solution to the
strong CP-problem and SUSY. The axino is the
fermionic partner of the axion. Similarly to the axion, the axino
couples to ordinary matter with a very tiny coupling proportional to
$1/\fa$ where $10^{9}-10^{10}\gev\lsim\fa\lsim10^{12}\gev$. 

It is plausible to consider the axino as the LSP since its mass is
basically a free parameter which can only be determined in specific
models.  As we have seen above, the neutralino has been accepted in
the literature as a ``canonical'' candidate for the 
LSP and dark matter. But with current LEP bounds between about 30 and
60~GeV (depending on a SUSY model), 
it becomes increasingly plausible that there
may well be another SUSY particle which will be lighter than the
neutralino, and therefore a candidate for the LSP and dark matter.

Primordial axinos decouple from the thermal soup very early, around
$T\simeq\fa$, similarly to the axions. The early study of Ragagopal,
\etal~\cite{rtw} concluded that, in order to satisfy $\Omega h^2<1$, the
primordial axinos had to be light ($\lsim 2$ keV), corresponding to
warm dark matter, unless inflation would be invoked to dilute their
abundance. In either case, one did not end up with axino as cold DM.

However, it has recently been shown~\cite{ckr} that the axino can be a
plausible {\em cold} dark matter candidate, and that its
relic density can naturally be of order the critical density. The
axino can be produced as a non-thermal relic in the decays of heavier
SUSY particles. Because its coupling is so much weaker, superparticles
first cascade decay to the next lightest SUSY partner (NLSP) for which
the most natural candidate would be the neutralino. The neutralino
then freezes out from thermal equilibrium at $T_f\simeq\mchi/20$. If
it were the LSP, its co-moving number density after freeze-out would
remain nearly constant. In the scenario of Covi \etal\ (CKR)~\cite{ckr}, the
neutralino, after decoupling from the thermal equilibrium, 
subsequently decays into the axino via, \eg, the process
\begin{equation}
\chi\ra\axino\gamma
\label{chitoagamma:eq}
\end{equation}
as shown in Fig.~1 in Ref.~\cite{ckr}. 
This process was already
considered early on in Ref.~\cite{kmn} (see also Ref.~\cite{rtw}) in the
limit of a photino NLSP and only for both the photino and axino masses
assumed to be very low, $\mphotino \leq 1 \gev$ and $\maxino\leq 300
\ev$, the former case now excluded by experiment. In that case, the
photino lifetime was typically much larger than 1 second thus normally
causing destruction of primordial deuterium from Big Bang
nucleosynthesis (BBN) by the energetic photon.
Avoiding this led to lower
bounds on the mass of the photino, as a function of $\fa$, in the
$\mev$ range~\cite{kmn}.

Because both the NLSP neutralino and the CKR axino are both 
massive (GeV mass range), the decay~(\ref{chitoagamma:eq}) is now typically
very fast. 
In the theoretically most favored case of a
nearly pure bino, \cite{chiasdm,kkrw}
the neutralino lifetime can be written as
\beq
\tau\simeq 3.3 \times 10^{-2} {\rm s}
\left(\frac{f_a/(N\cayy)}{10^{11}\gev}\right)^2
\left(\frac{100\gev}{\mchi}\right)^3
\label{binolife:eq}
\eeq
where the factor $N\cayy$ is of order one.
One can see that it is not difficult to ensure that the decay takes
place well before 1 second in order for avoid 
problems with destroying successful predictions of Big Bang
nucleosynthesis.  The axino number density is equal to that of the
NLSP neutralino. Therefore its relic abundance is 
$\abunda= \left(\maxino/\mchi\right)\abundchi$.
The axinos are initially relativistic but, by the time of matter
dominance they become red-shifted by the expansion and become cold DM.

There are other possible production mechanisms and cosmological
scenarios for massive axinos.\footnote{For a recent comprehensive
  study, see~\cite{ckkr}.} Even if the primordial population of
axinos is inflated away (which would happen if the reheating
temperature $T_{\rm reh}\ll\fa$), they can be regenerated from thermal
background processes at high enough $T_{\rm reh}$.

\section{Conclusions}
Looking for the invisible is not easy but is certainly
worthwhile. The discovery of dark matter will not only resolve
the mystery of its nature but is also likely 
to provide us with much information
about the particle physics beyond the Standard Model. What I find most
encouraging is that we have a very good chance of fully testing the
neutralino as a WIMP by the end of the decade. Detecting other
candidates, like the axino, will be the task for the more distant future.

\vspace*{0.5cm}
\paragraph
{\em Note Added:} After the Conference both the DAMA and CDMS
Collaborations published new results. Based on the combined statistics
of $57,986\kgday$ of data collected in a NaI detector since November
'96, the DAMA Collaboration reported~\cite{dama2000} a
statistically significant ($4\,\sigma$~CL) effect which it interprets
as being caused by a WIMP annual modulation signal.

The CDMS experiment using germanium and silicon
crystals at Stanford published~\cite{cdms-2000} a new
limit on scalar WIMP-proton cross section. The new result is based on
the total of $10.6\kgday$ of data collected a current shallow site
(17~mwe) at Stanford during 1999. A powerful event-by-event
discrimination method allows CDMS to reach a sensitivity matching that of
DAMA with only less than $0.2\%$ of DAMA's statistics. 

The new $90\%$~CL CDMS limit excludes most of the signal region
claimed by DAMA. In particular, it fully rules out the previous
$2\,\sigma$ region based on the combined data of $19,511\kgday$ from
runs~I and~II~\cite{dama98two} and the new $3\,\sigma$ region at more
than $84\%$~CL.

An updated compilation of these and other data can be found on
the Web: http://cdms.berkeley.edu/limitplots/.)

\end{document}